\documentclass[sn-mathphys,Numbered]{sn-jnl}

\usepackage{multirow}%
\usepackage{amsmath,amssymb,amsfonts}%
\usepackage{amsthm}%
\usepackage{mathrsfs}%
\usepackage[title]{appendix}%
\usepackage{xcolor}%
\usepackage{textcomp}%
\usepackage{manyfoot}%
\usepackage{booktabs}%
\usepackage{algorithm}%
\usepackage{algorithmicx}%
\usepackage{algpseudocode}%
\usepackage{listings}%

\usepackage{graphicx, subfigure} 

\usepackage[normalem]{ulem} 
\usepackage{siunitx}
\usepackage{hyperref}
 
\usepackage{epsfig}

\begin{document}

\title{Cooling of a granular gas mixture in microgravity}

\author*[1,2,4]{\fnm{Dmitry} \sur{Puzyrev}}\email{dmitry.puzyrev@ovgu.de}

\author[1,2,3,4]{\fnm{Torsten} \sur{Trittel}}

\author[2,3,4]{\fnm{Kirsten} \sur{Harth}}

\author[2,3,5]{\fnm{Ralf} \sur{Stannarius}}

\affil[1]{\orgdiv{Department of Nonlinear Phenomena,
Institute of Physics}, \orgname{Otto von Guericke University Magdeburg}, \orgaddress{\street{Universit\"atsplatz 2}, \city{Magdeburg}, \postcode{39106}, \country{Germany}}}

\affil[2]{\orgdiv{Research Group ‘Magdeburger Arbeitsgemeinschaft f\"ur Forschung unter Raumfahrt-und
Schwerelosigkeitsbedingungen’ (MARS)}, \orgname{Otto von Guericke University Magdeburg}, \orgaddress{\street{Universit\"atsplatz 2}, \city{Magdeburg}, \postcode{39106}, \country{Germany}}}

\affil[3]{\orgdiv{Department of Engineering}, \orgname{Brandenburg University of Applied Sciences}, \orgaddress{\street{Magdeburger Str. 50}, \city{Brandenburg an der Havel}, \postcode{14770}, \country{Germany}}}

\affil[4]{\orgdiv{Department of Microgravity and Translational Regenerative Medicine, Medical Faculty}, \orgname{Otto von Guericke University Magdeburg}, \orgaddress{\street{Universit\"atsplatz 2}, \city{Magdeburg}, \postcode{39106}, \country{Germany}}}

\affil[5]{\orgdiv{Institute of Physics}, \orgname{Otto von Guericke University Magdeburg}, \orgaddress{\street{Universit\"atsplatz 2}, \city{Magdeburg}, \postcode{39106}, \country{Germany}}}


\abstract{
Granular gases are fascinating non-equilibrium systems with interesting features such as spontaneous clustering and non-Gaussian velocity distributions. Mixtures of different components represent a much more natural composition than monodisperse ensembles, but attracted comparably little attention so far. We present the first experimental observation and characterization of a mixture of rod-like particles with different sizes and masses in microgravity. Kinetic energy decay rates during granular cooling and collision rates were determined and Haff's law for homogeneous granular cooling was confirmed.  Thereby, energy equipartition between the mixture components and between individual degrees of freedom is violated. Heavier particles keep a slightly higher average kinetic energy than lighter ones. Experimental results are supported by numerical simulations.
}

\maketitle
\section{Introduction}
Exploring the behavior of granular gases in microgravity environments holds an immense scientific and practical significance. This area of research advances our understanding of physics, engineering, and even space exploration. From the viewpoint of fundamental physics, ensembles of individual macroscopic particles colliding in a manner similar to gas molecules offer unique perspectives on the basic laws of multiparticle physics. Microgravity allows to observe and analyze pure granular interactions, eliminating the complex influence of gravity. This can lead to breakthroughs in understanding particle dynamics, energy dissipation, and entropy production.

In astronomy and cosmology, understanding behavior of granular systems sheds light on the formation and dynamics of celestial bodies, such as asteroids, comets, planetesimals and planetary rings. One can learn a lot about the way these objects evolve and interact. The study of granular gases can also provide valuable insights into energy dissipation as well as energy and heat transfer mechanisms, with manifold implications for the design of efficient applications on Earth and in space.
It should be emphasized that microgravity experiments with granular gases can serve as captivating educational tools, inspiring students in physics, space science and computer vision. 

Whereas the majority of previous experiments and theoretical investigations of granular gases was focused on monodisperse systems, more realistic studies have to take into account that such systems in general are polydisperse. Mixtures introduce an additional layer of complexity. One of the fundamental questions is the partition of kinetic energies among the constituents, and among their different degrees of freedom. Additionally, one may investigate the emergent behaviors and self-organization in long-term experiments with these mixtures.

This study is one of the first steps towards a generalization of granular gases to polydisperse systems. We report experiments and numerical simulations of bidisperse mixture and compare energy partition, dissipation (granular cooling) and collision statistics observed in microgravity experiments and numerical simulations.

The dissipative
character of particle interactions determines the ensemble
properties of granular gases: Clustering \cite{Goldhirsch1993,Kudrolli1997,Falcon1999,Falcon2006,Opsomer2011,Sapozhnikov2003,Olafsen1998,Puzyrev2020,Puzyrev2021,PuzyrevDmitry2021}, non-Gaussian velocity
distributions \cite{Olafsen1998,Olafsen1999,Nichol2012,Kudrolli2000,Rouyer2000,Losert1999,Huan2004,Hou2008,Aranson2002,Kohlstedt2005,Tatsumi2009,Schmick2008}, and anomalous pressure scaling \cite{Falcon2006,Geminard2004} were described. In contrast to a large number of numerical simulations dealing with these systems, there is a comparably small experimental basis, mostly restricted to two-dimensional (2D) systems, e.~g. Refs. \cite{Tatsumi2009,Hou2008,Grasselli2009,Yanpei2011,Kohlstedt2005,Aranson2002,Maass2008}. Few 3D experiments in microgravity have been reported, with spherical grains \cite{Falcon1999,Falcon2006,Yu2020},
ellipsoids \cite{Pitikaris2022} and rods \cite{Harth2013, Harth2018}.
Most experiments were performed with monodisperse systems, providing fundamental insights into the relations between microscopic processes (particle collisions) and ensemble characteristics (granular temperatures and spatial homogeneity), but they lack a typical feature of most of the natural granular gases, viz. the composition of differently shaped and sized constituents.
We report experiments with a 3D mixture of rods of same lengths but with two different diameters. 
Statistical data were extracted in the initial heating phase and during free cooling.

Already in the 1980s, Peter Haff \cite{Haff1983} proposed a scaling law for the kinetic energy loss of a homogeneously cooling dense granular gas of frictionless monodisperse spheres. He predicted a time dependence of the form 
\begin{equation}
\label{eq:Haff}
 {E_{\rm kin}(t)}=\frac{E_0}{(1+t/\tau_{\rm H})^2},
\end{equation}
which yields the scaling $E_{\rm kin}\propto t^{-2}$ for times $t\gg \tau_{\rm H}$. $E_0$ is the kinetic energy at time $t=0$. The Haff time $\tau_{\rm H}(E_0)$ for a given initial state
defines a time scale of energy loss. It depends on material properties and the shape of the grains, and on other system parameters.

Haff's law was confirmed for free cooling dilute ensembles of monodisperse rod-like particles \cite{Harth2018},  spheres~\cite{Yu2020} and oblate ellipsoids~\cite{Pitikaris2022}. Estimates for the mean kinetic energies and $\tau_{\rm H}$ were obtained. For mixtures, an important and unresolved question is whether and how the components differ in their kinetic and cooling parameters. 
Granular gas mixtures of spherical particles with different radii and masses were analysed in numerical simulations by Bodrova et al. \cite{Bodrova2014, Bodrova2020}. Heating and cooling were considered, yet particle rotations were disregarded for simplicity. Larger, heavier particles were found to have a higher granular temperature than lighter ones. The particular distribution of kinetic energies depends on the choice of the contact model in the simulation. Different granular temperatures for different components are quite common in granular matter, they were, e.g., also predicted for vibrated granular fluids \cite{Brey2011}.
Experimental data for a confirmation of these theoretical predictions were not available so far.

\begin{figure}[htbp]
\includegraphics[width=1.0\linewidth]{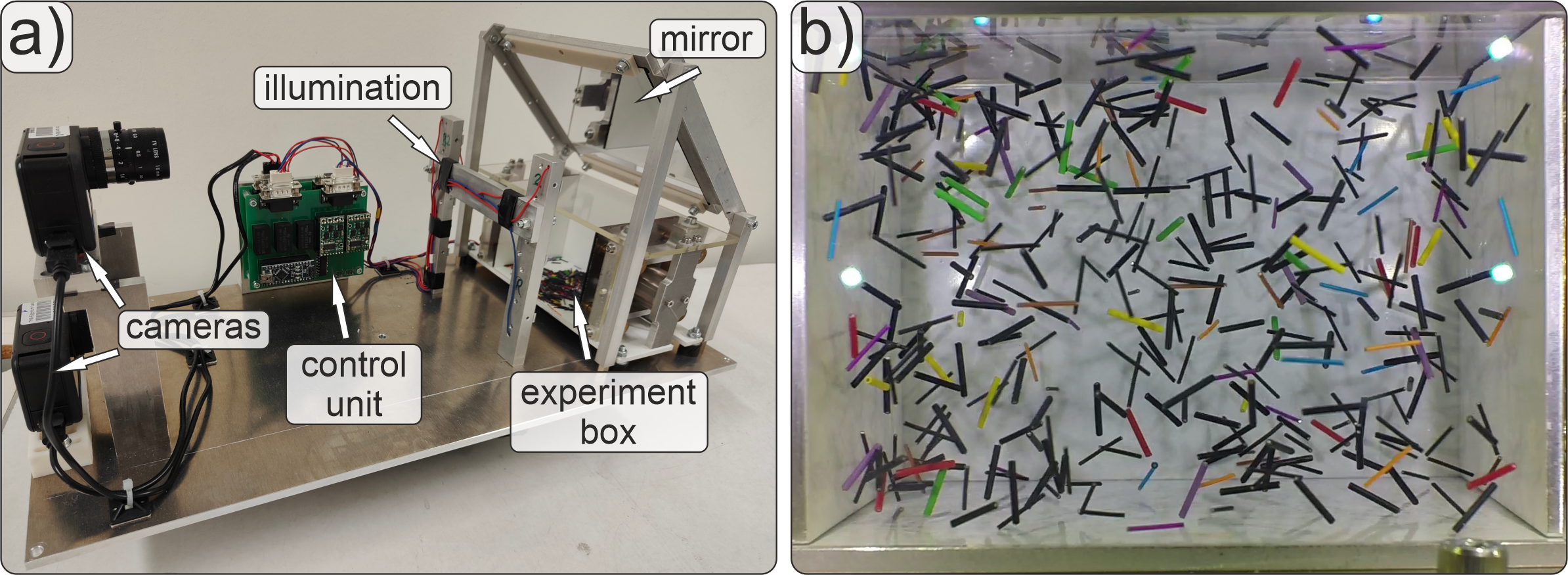}
\caption{a) Experimental setup: Inside the container (experiment box) is a mixture of elongated grains which can be excited by two vibrating walls. It is observed by two cameras though front and top transparent walls. b) Example of video frame which shows granular gas mixture during its cooling in microgravity. The image width is about 12 cm.}
\label{fig:Setup}
\end{figure}

\section{Experimental setup and parameters}
In the present paper, we analyze the dynamical properties of a two-component granular gas mixture.  
Our experimental setup follows the description given in Ref.~\cite{Harth2018} and is presented on Figure~\ref{fig:Setup}: Two cameras viewing along axes $y$ and $z$ were used for a stereoscopic observation of a granular gas consisting of 384 rods (192 of each of sort) in a container with dimensions $L_x \times L_y \times L_z =  11.2 \times 8.0 \times 8.0$ cm$^3$. The rods consist of insulated copper wire pieces of length $\ell=10$ \si{mm} and diameters $d_1=0.75$ mm (component 1) and $d_2=1.35$ mm (component 2), respectively. The mass of the thin rods is $m_1=22$ mg, their moment of inertia for rotations about the long axis is $J_{\parallel 1}=0.99$ \si{\pico\newton.\meter.\second^2} and perpendicular to it $J_{\bot 1}=183$ \si{\pico\newton.\meter.\second^2}. For the thicker particles, the mass is $m_2=37.5$ \si{mg}, 
$J_{\parallel 2}=4.6$ \si{\pico\newton.\meter.\second^2} and $J_{\bot 2}=315$ \si{\pico\newton.\meter.\second^2}. For easier detection and tracking, 48 rods of each type were colored, the rest serves as ``thermal'' background. The experiment was performed in the ZARM drop tower in Bremen, where microgravity ($\mu g$) is achieved for about 9.2 seconds. 
Initially, the system was excited by sinusoidal vibration of two side walls of the container (in $x$ direction) at an amplitude of $A=0.24$ cm and frequency $f = 30$ Hz, i.~e. a maximum plate acceleration of $\approx 8~g$. Then, vibration was stopped and the granular gas mixture was left without energy input in a granular cooling phase.

3D particle detection in the experiments, tracking and trajectory post-processing mainly follow the outlines of Refs.~\cite{Puzyrev2020,Harth2018}.
The rod mixture is excited during the first two seconds of $\mu g$, then it undergoes granular cooling.

\section{Numerical simulations}
In order to support the experimental analysis, we performed a numerical simulation. It provides the opportunity to extract particular properties that are not accessible in the experiment. At the same time, comparison of the simulation results with the experiment helps to choose a suitable collision model and realistic material parameters.
We used a hybrid GPU-CPU implementation of discrete element modelling (DEM) \cite{RubioLargo2016,Rubio2015,heaping_fischer,Pongo2021,Pongo2022}, adapted to systems with moving walls. The collision detection and Hertz-Mindlin contact force model follows the previous simulations with spherocylinders \cite{Pongo2021,Puzyrev2021,Pongo2022}. In addition, our program was modified to simulate the bidisperse mixture of particles with the same length and diameters as in the experiment.
The energy loss was quantified using an effective restitution coefficient $e_n=0.7$ and friction coefficient
$\mu=0.4$. These values  provide reasonable agreement with the experiment, even though the computed Haff times are somewhat larger. The Young modulus was set to $Y = 5$ GPa, yet its particular value does not influence the results noticeably as long as it is large enough to restrict excessive particle overlap, and small enough so that the collisions are resolved in a sufficient number of simulation steps ($\delta t = 5\times10^{-8}$s) \cite{Antypov2011}.
Container and excitation parameters were the same as in the experiment. 

\section{Results and discussion}
\subsection{Cooling rate and Haff time}

\begin{figure}[htbp]
\includegraphics[width=1.0\linewidth]{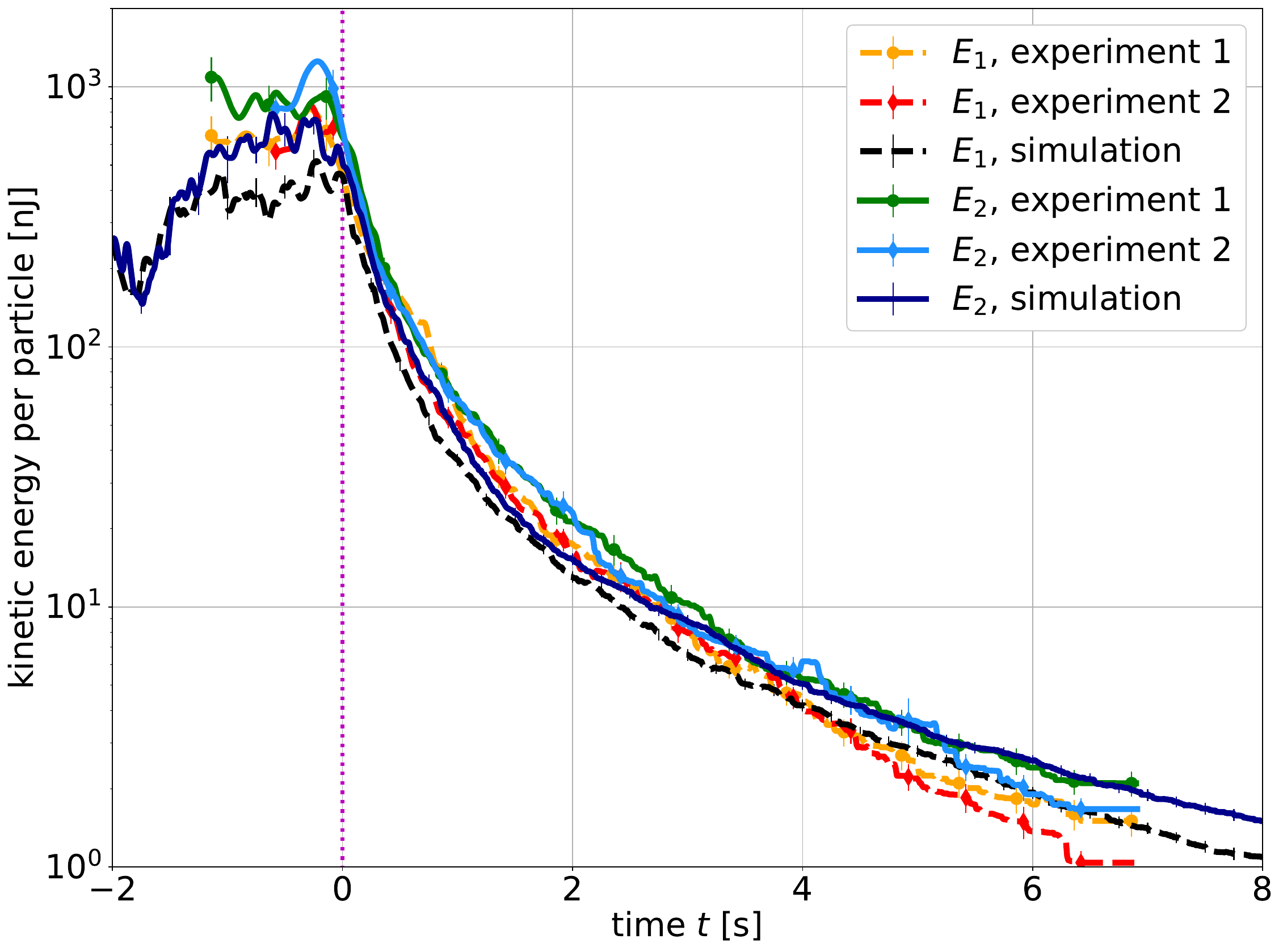}
\caption{Total kinetic energy for the two mixture components from experiment and simulation in comparison. 
 Time refers to the stop of excitation (vertical dashed line), 2 s after entry into the microgravity phase. }
\label{fig:EnTot}
\end{figure}

\begin{figure}[htbp]
\includegraphics[width=1.0\linewidth]{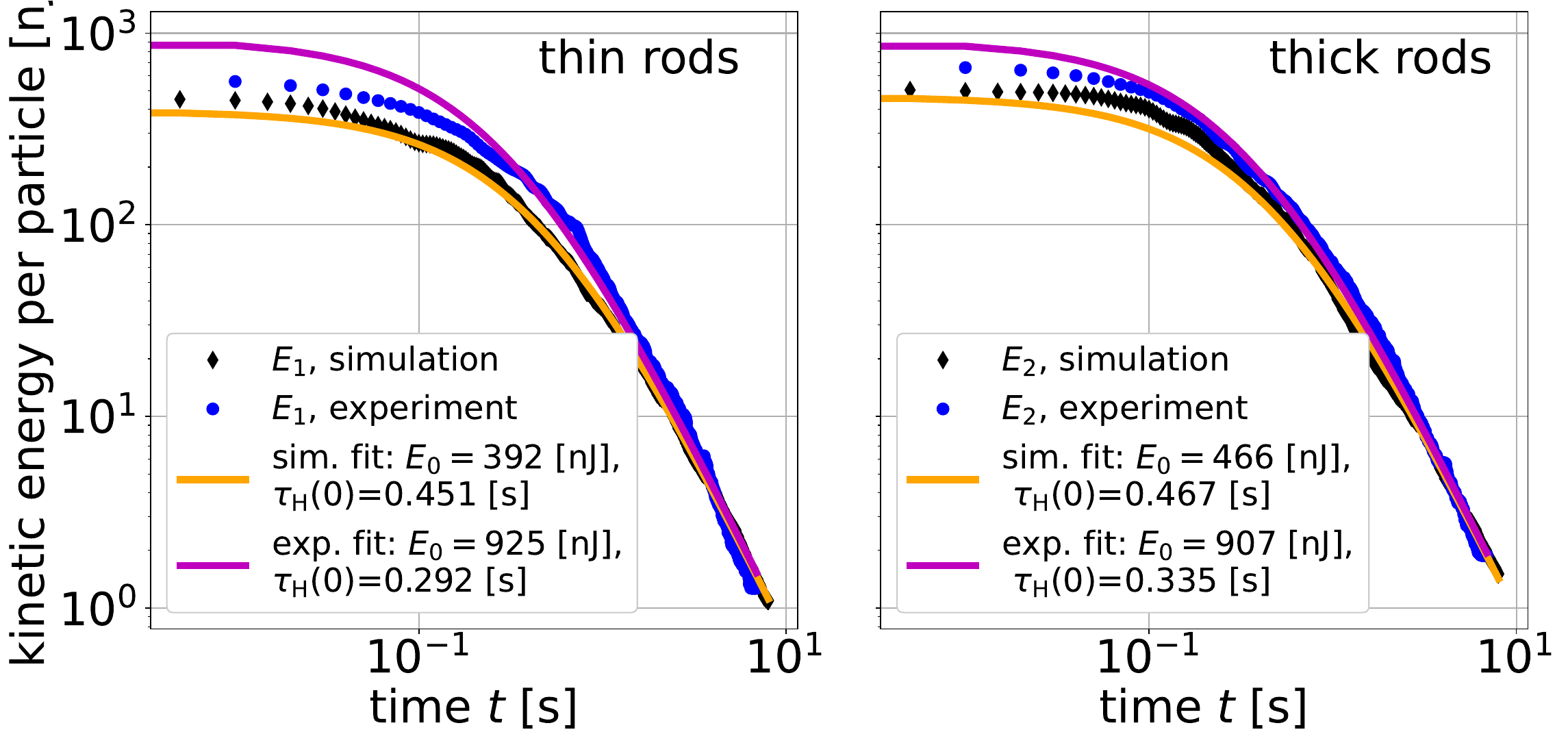}
\caption{Total kinetic energy for the two mixture components in double logarithmic scale. Both are fitted well for $t>0.2$~s with Eq.~\eqref{eq:Haff} and the parameters given in the graphs.
}
\label{fig:EnTotHaff}
\end{figure}

Figure~\ref{fig:EnTot} shows the decay of the average kinetic energy per particle separately for the two components, thin ($E_1$) and thick ($E_2$) rods. Time is counted from the start of granular cooling after the excitation was switched off. Vertical lines represent the uncertainty which arises mainly from the limited ensemble size of evaluated rods \cite{Harth2018}. From the experimental data, two separate runs (drops) are presented. One observes satisfactory agreement between both experimental runs and simulation
\footnote{Note that in the experiment, rotations about the long rod axis were not observable
and are thus not included in the evaluation.}. In the subsequent figures, the data for both experimental runs are combined.
The mean energies per particle at the end of the heating phase are  $E_1\approx 580$~\si{\nano\joule} and $E_2\approx 660$ \si{\nano\joule} (see Fig.~\ref{fig:EnTot}). 
Figure~\ref{fig:EnTotHaff} shows that Haff's law, Eq.~(\ref{eq:Haff}),  well reproduces the loss of kinetic energy during granular cooling for both experimental and simulated data, except for the initial 0.2~s after heating was switched off. There, it fails because the heated state is too inhomogeneous \cite{Harth2018}, and this is the reason why the fit parameter $E_0$ does not represent the actual initial energy.
Most importantly, fitted Haff times $\tau_{\rm H}$ differ only slightly for the two components, namely, $\tau_{\rm H1} = 0.45\pm 0.02$ \si{\second} and $\tau_{\rm H2} = 0.47\pm 0.02$ \si{\second} in the simulation, while $\tau_{\rm H1} = {0.29 \pm 0.03}$ \si{\second} and $\tau_{\rm H2} = 0.34\pm 0.03$ \si{\second} in the experiment.
Within the statistical accuracy of our data, this is consistent with
equal Haff times for both components. 
If a slight systematic deviation actually exists, it means that the heavier particles initially cool slower until an equilibrium is reached with equal Haff times and the energy distribution at later stages is slightly shifted further away from equipartition. 
The reported difference of Haff times between experiments and simulation is not problematic. Note that the Haff time has the property $\tau_{\rm H}(t_0+t) = \tau_{\rm H}(t_0)+t$, so that the difference of the initial Haff times in experiment and simulation appears as a simple time shift of 0.15 s between the experimental and simulated data (cf. Fig.~\ref{fig:EnTot}).

\subsection{Energy partition}

\begin{figure}[htbp]

\includegraphics[width=1.0\linewidth]{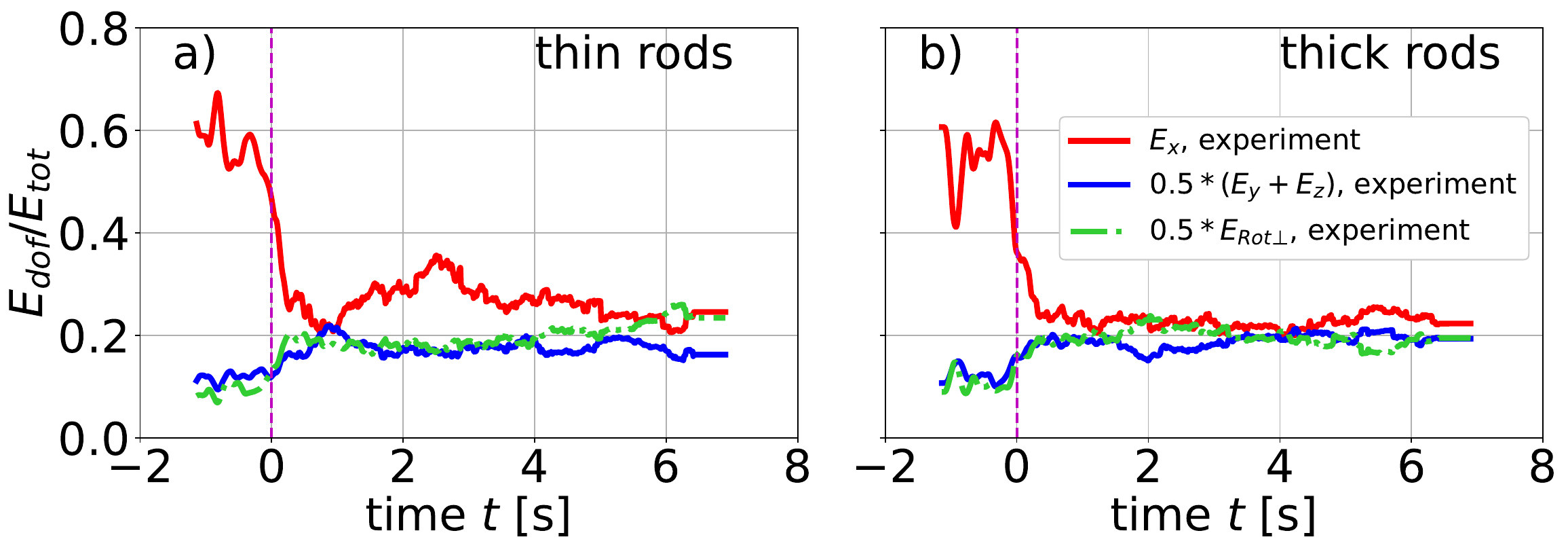}
\includegraphics[width=1.0\linewidth]{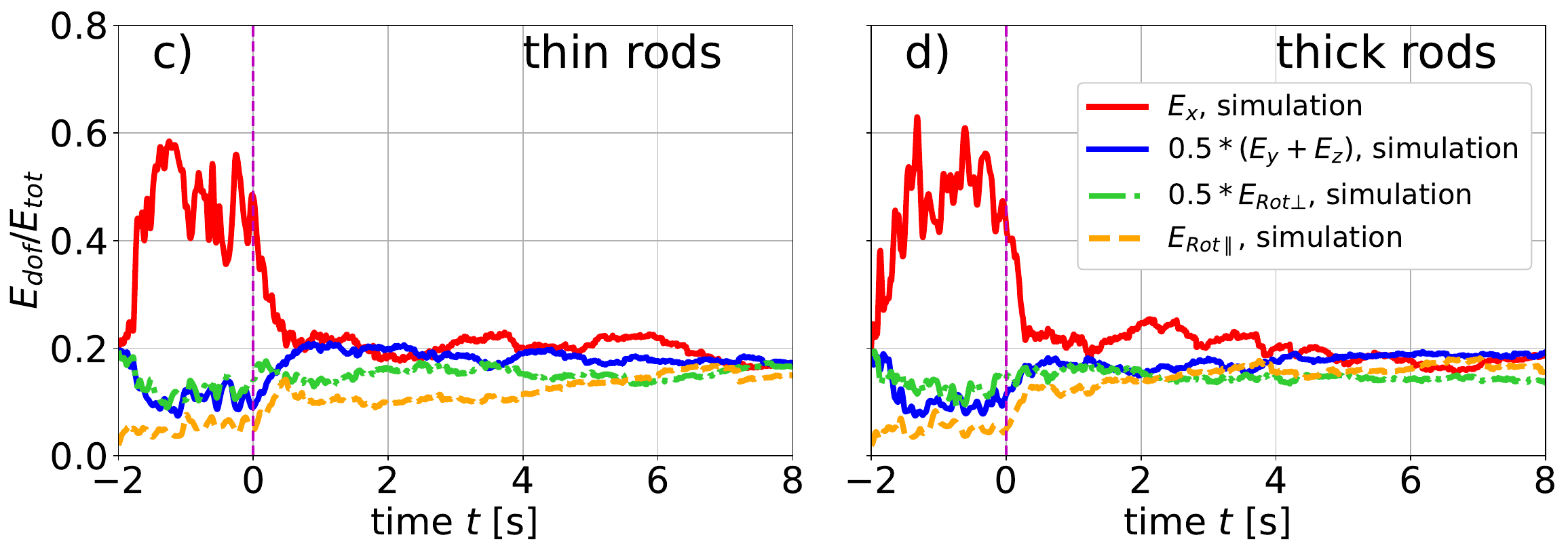}

\caption{Partition of the mean particle energies per DOF, (a,b) experiment, (c,d) simulation. Here and in following figures, vertical dashed lines mark the start of cooling.}
\label{fig:Epart}
\end{figure}

We first compare the shares of kinetic energies per individual degree of freedom (DOF) of each rod type.
An excess of translational energy along the `active' direction $x$ is observed for both components during excitation (Fig.~\ref{fig:Epart}). The average energy for the directly excited DOF can reach more than half of the total kinetic energy. This is in accordance with previous findings \cite{Huan2004,Costantini2005,Feitosa2004,Yanpei2011,Harth2013,Pongo2021} and arises mainly from `hot' particles directly after collisions with vibrating walls. In addition, we find a granular temperature gradient towards the center of the container in the heated $x$-direction, similar to experiments with excited dense granular ensembles \cite{Noirhomme2017}. The velocity distribution functions are non-Gaussian. 
The continuously heated phase was not the primary focus of this study. A thorough analysis of energy balance in continuously heated systems deserves a more dedicated experiment.

Figure \ref{fig:Epart} shows the energy partition between DOF associated with translational motion ($E_x,E_y,E_z$) as well as two rotational DOF about the short rod axes ($E_{\rm Rot\bot}$). 
After the onset of cooling, the share of energy associated with translation along $x$ rapidly decreases.
We obtain kinetic energies per DOF which are close to equipartition, except for slight residual dominance of translations along $x$.
Energy partition for cooling monodisperse rod ensembles was studied in Refs.~\cite{Harth2018, Harth2013}: After a relatively short initial period, the particles were found to reach a steady partition of $E_x,E_y,E_z$ and $E_{\rm Rot\bot}$. The average energy associated with the rotational DOF was found around $10{\text -}20\%$ less than for translations. 

The kinetic energy $E_{\rm Rot\parallel}$ contained in the 6$^{th}$ degree of freedom, rotation around the long rod axis, is quite difficult to determine experimentally, and was not accessible in the present experiments. An approximate value was given in Ref.~\cite{Harth2013} where it was estimated that $E_{\rm Rot\parallel}$ is about one order of magnitude lower than the mean energies of the other DOF.
In our simulations, all DOF are directly accessible. The energy partition extracted from the simulation (Fig.~\ref{fig:Epart}~c,d) yields energy levels for $E_{\rm Rot\parallel}$ contributing around $10\%$ to the total kinetic energy at the begin of cooling, and even slightly growing afterwards. This is significantly more than the value obtained earlier experimentally \cite{Harth2013}. 
These rotations about the long axis are exclusively excited by frictional contacts. Trusting the experiment, we presume that the
realization of these frictional contacts in the simulation is
not yet satisfactory and needs refinement.
\begin{figure}[htbp]
\includegraphics[width=0.8\linewidth]{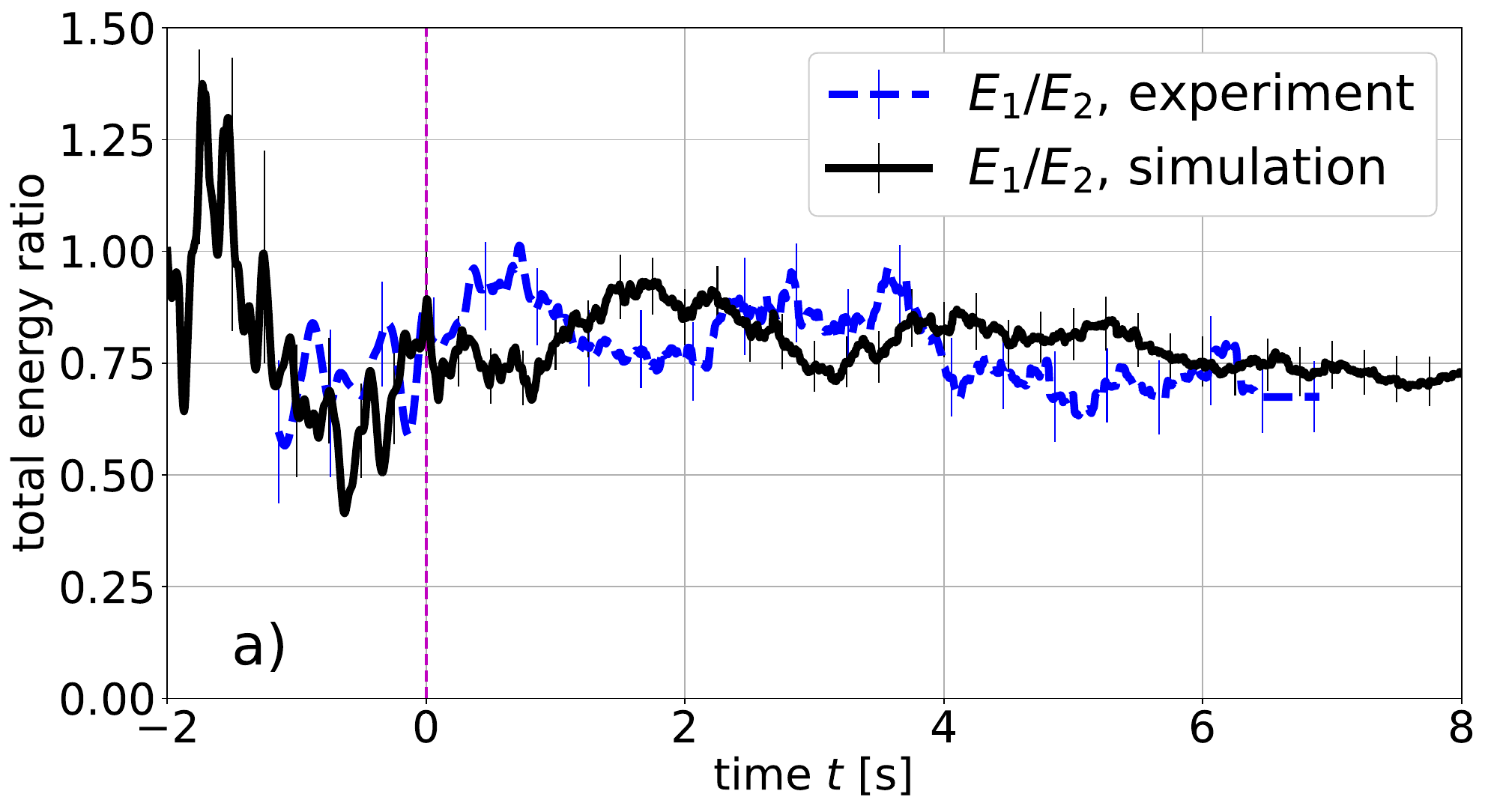}\\
\includegraphics[width=0.8\linewidth]{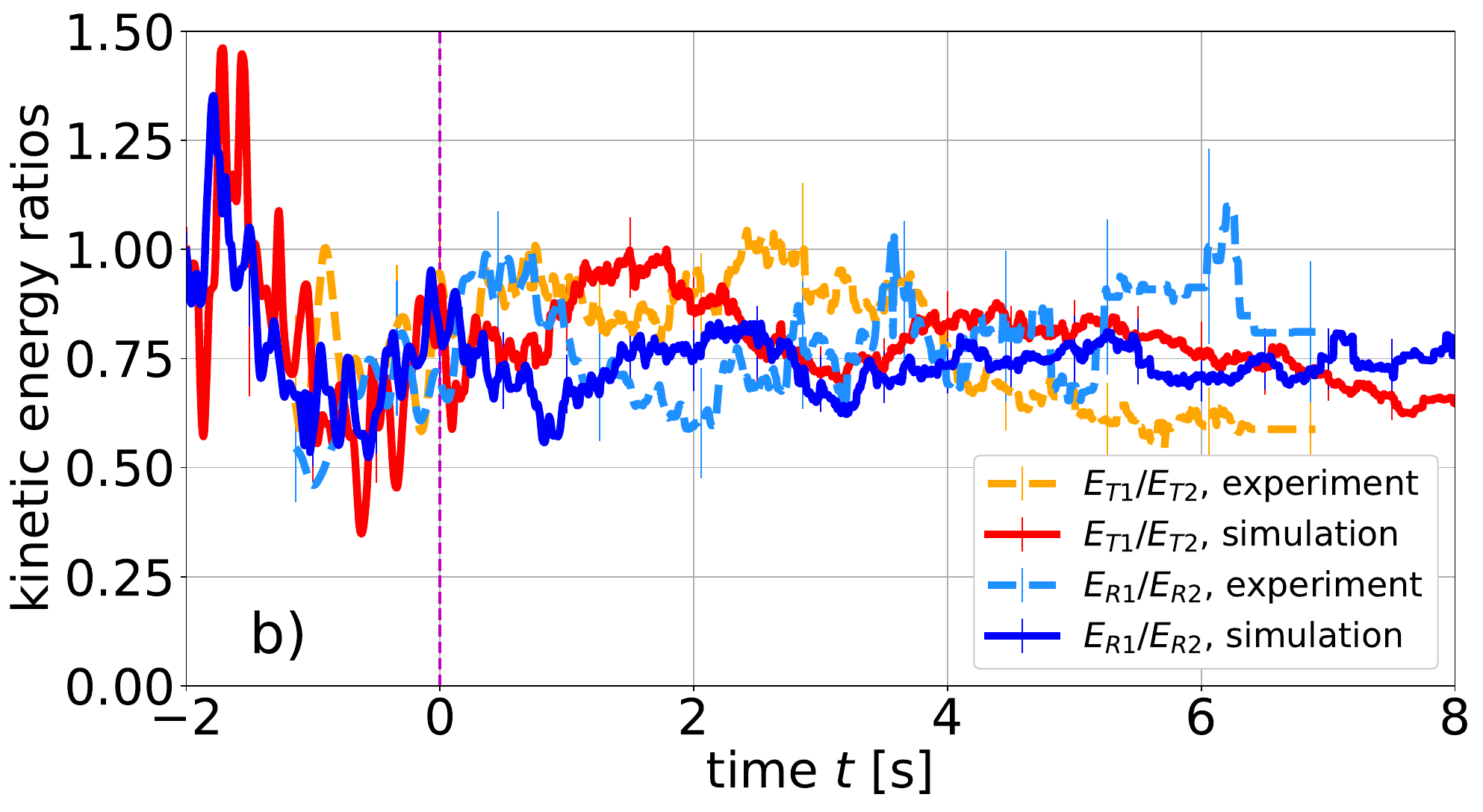}
\caption{a) Ratio of the total kinetic energies $E_1,E_2$ per particle for the mixture components 1 and 2.
The red dashed line marks the start of cooling.
b) Ratios of translational ($E_{T1},E_{T2}$) and rotational ($E_{R1},E_{R2}$) kinetic energies per particle for the mixture components.}
\label{fig:EnRatios}
\end{figure}

A crucial question for granular gas mixtures is the dependence of the mean kinetic energy on particle properties such as relative particle size and mass. In our experiments, the mass ratio is $m_1/m_2 = 0.59$ (diameter ratio $d_1/d_2 = 0.56$). 
An excess of average kinetic energy of the larger particles persists during the complete cooling phase, as seen from the ratio $E_1/E_2 \approx 0.8$ of the average total (observed) kinetic energies per particle for the mixture components in Fig.~\ref{fig:EnRatios}(a), both during the heating and cooling stages.

Figure~\ref{fig:EnRatios}(b) shows separately the ratios of translational ($E_{T1}/E_{T2}$) and rotational ($E_{R1}/E_{R2}$) energies. The data are more noisy than for the total energy due to the permanent energy exchange between translational and rotational DOF. Nevertheless, we observe satisfactory statistical agreement between experiment and simulation.
\clearpage

\subsection{Collision statistics}

\begin{figure}[htbp]
\includegraphics[width=1.0\linewidth]{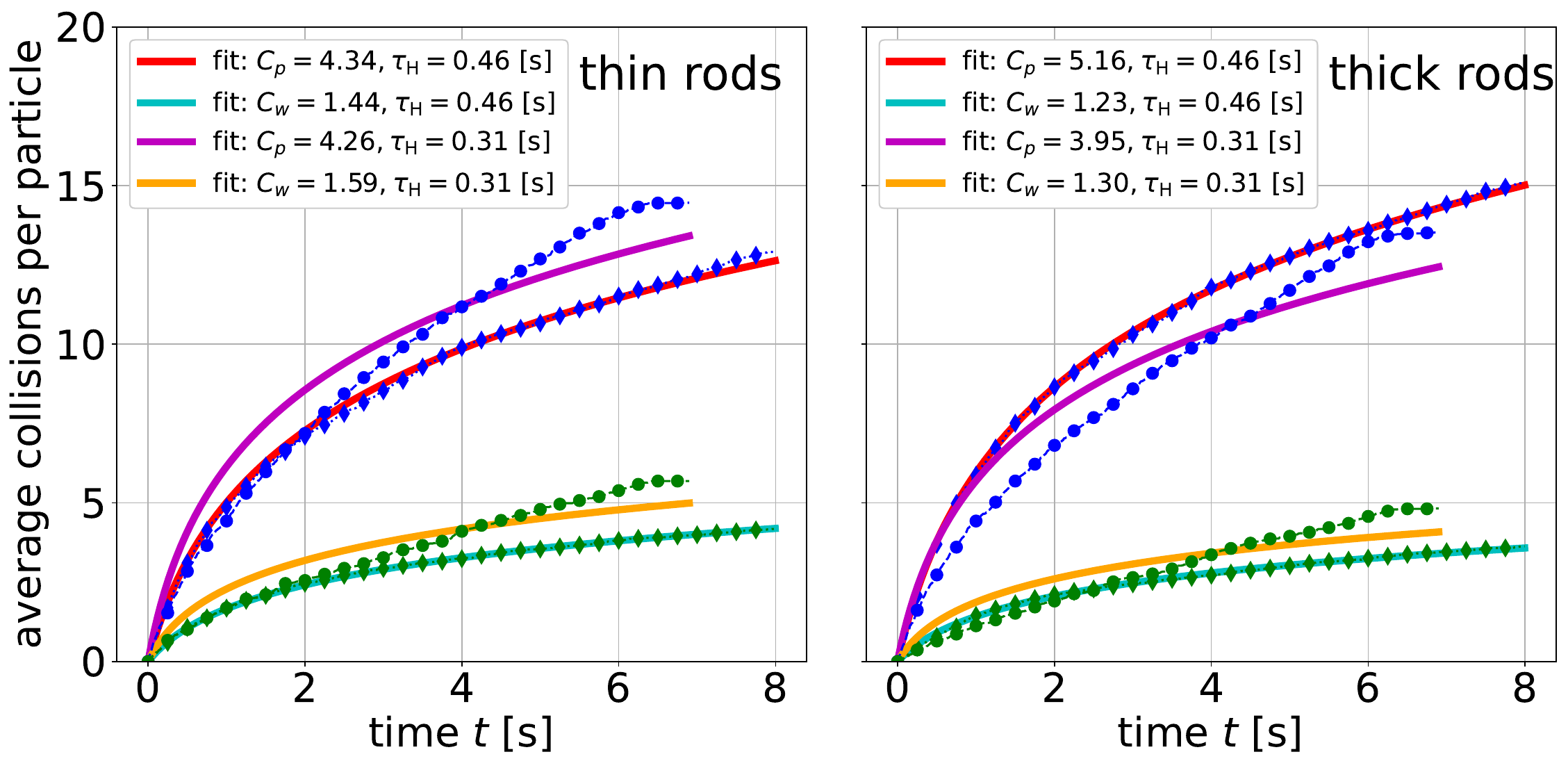}
\caption{Cumulative statistics of particle-particle (blue markers) and particle-wall collisions (green markers). Solid lines are fits with logarithmic functions of the form given in Eq.~(\ref{eq:coll_fit}). Double collisions of the rods with a wall  \cite{Trittel2017} count as one.
}
\label{fig:Collisions}
\end{figure}

\begin{figure}[htbp]
\includegraphics[width=1.0\linewidth]{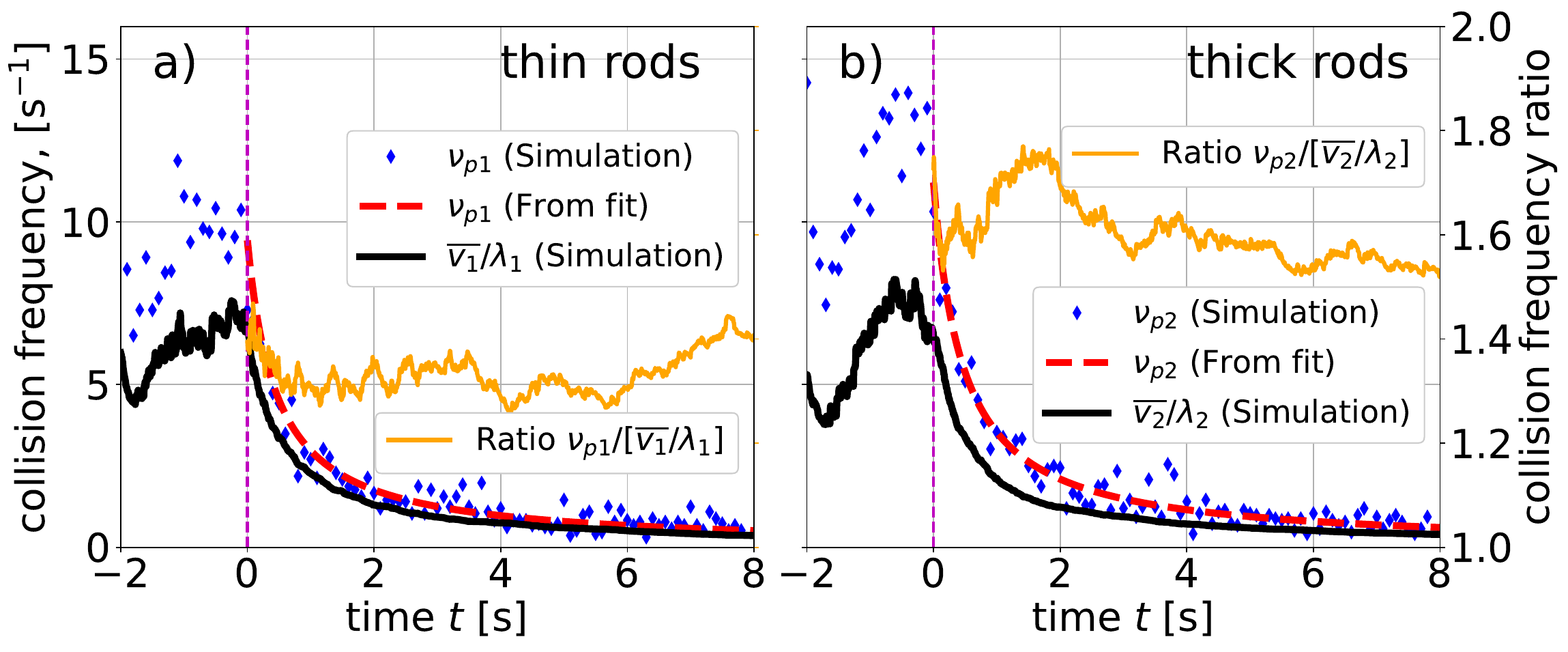}
\caption{Average particle-particle collision frequencies $\nu_{p1,2}$. Orange line shows the ratio of collision frequency observed in simulation to one obtained from mean velocity and theoretical mean free path estimate.}
\label{fig:Coll_freqs}
\end{figure}

The statistics of collisions as the elementary steps of the cooling process are compared in Figures \ref{fig:Collisions} and \ref{fig:Coll_freqs}.
Figure \ref{fig:Collisions} shows the cumulative number of collisions of each particle type with other particles and with the container walls, extracted directly from the simulations, as well as from an analysis of experimental particle trajectories. 
Following Haff's model, the average cumulative number of particle-particle collisions in the system can be approximated by \cite{Costantini2005,Villemot2012,Harth2018}
\begin{equation}
\label{eq:coll_fit}
N_{\rm C}(t) = C_p \ln
\left(1+\frac{t}{\tau_{\rm H}}\right) ,
\end{equation}
where time $t$ starts with the beginning of cooling. 
Here, $C_p$ is a positive constant related to the mean energy loss in a single particle-particle collision \cite{Harth2018,Villemot2012}. Since the number of collisions is the ratio of the mean speed of the particles and some time-independent geometric parameter (i.e. the mean free path) both particle-particle and particle-wall collisions should obey the same logarithmic dependence as in Eq.~(\ref{eq:coll_fit}), with different prefactors $C_p$ and $C_w$.
The four curves in Figure~\ref{fig:Collisions} were fitted by Eq.~\eqref{eq:coll_fit} with mean initial Haff times $\tau_{\rm H}=0.31$~s (experiment) and 0.46~s (simulation). While the simulation collision numbers are accurate and match the fits quite nicely, the experimental results differ significantly. 
The possible reason is that not all collisions could be correctly identified in the videos with our automated collision detection, both in the initial phase where particles move too fast and in the later stages of cooling where noise in the detected trajectories can be mistaken for collisions. A more accurate collision detection would be possible in experiments with somewhat lower packing fractions and an improved camera setup. 
Nevertheless, it is evident that particle-wall collisions are around $2.5-3$ times less frequent than particle-particle ones for the thinner rods and around $3.5-4$ times less frequent for the thicker rods.  

For a comparison, we modified the formula for the collision cross-section of rods \cite{Harth2018} to include cylinders of different diameters. A random particle orientation respective to the flight direction and a uniform, homogeneously mixed particle distribution are assumed for simplicity. The number of rods of each type be $N_1$ and $N_2$, resp. Then, we obtain estimates of the mean free paths $\lambda_{1}$ and $\lambda_{2}$ as: 
\begin{equation}
\label{eq:FreePath}
\lambda_i = \frac{V}{\sqrt{2}}\left( \frac{1}{N_1 \sigma_{1i} +N_2\sigma_{2i}}\right),\hspace{0.5cm}(i,j=1,2,) ,
\end{equation}
where the scattering cross sections $\sigma_{ij}$ corresponding to collisions of particle types $i$ and $j$ can be approximated as:
\begin{equation}
\label{eq:CollisionAreas}
\sigma_{ij} = \frac{\pi \ell^2}{8}
+\left(\frac{3+\pi}{4}\right) \left(d_i+d_j\right)\ell
+\frac{9 d_i d_j }{2\pi}
+\frac{d_i^2+d_j^2}{2}.
\end{equation}
The estimated mean free paths are $\lambda_{1}=1.92$~cm and $\lambda_{2}=1.66$~cm, resp., in absence of wall collisions. A rough estimate of the mean free path of a sphere in a cubic container with sides $\ell$ yields $\lambda_{\rm w}\approx 0.583~\ell$, which for our experiment is roughly $2.7~\lambda_1$ or $ 3.2~\lambda_2$. Thus, the predicted ratio of collisions with particles and with walls should be around 3, in fairly satisfactory agreement with the experiment. 

Another detail of the collision statistics is elucidated in Fig.~\ref{fig:Coll_freqs}. When the particle-particle collisions counted in the simulation are compared to the expected value from $\bar v/\lambda$ with the mean particle velocity $\bar v_{1,2}$ and the mean free paths $\lambda_{1,2}$ from Eqs.~(\ref{eq:FreePath},\ref{eq:CollisionAreas}), a factor of $\approx 1.5$ is evident. The mean free paths directly extracted from the simulations are also $\approx 1.5$ times shorter than their theoretical estimates. A possible reason is that the simplified collision model considers only non-rotating rods. Fast rotations about the short axes obviously increase the scattering cross sections.

\section{Summary}
Summarizing, a first quantitative confirmation of the main features of the granular cooling of a bidisperse mixture was achieved here. Heavier particles have larger mean energies both during excitation and during cooling. The ratios $E_1/E_2$ of average kinetic energies per particle were found to be around $0.8$ both in experiment and simulation (with the mass ratio being $0.59$ and the moment of inertia ratio $0.58$ for the rotations about the short axes). The Haff time determined for heavier particles was found to be slightly larger than that of the lighter ones, which might indicate a slower cooling of the heavier particles until an equilibrium $E_1/E_2$ is reached, that is smaller than the value reached within our observation time scale. In perspective, this study shall be extended to bidisperse mixtures of particles which are significantly different to each other \cite{Hidalgo2016}, particularly with different shapes. 
The investigation of polydisperse mixtures is highly desired, but several problems including the automatic detection and distinction of components need to be solved first.

\section*{Methods}
The ensemble was observed with two video cameras {\em GoPro Hero 3 Ribcage}. The image resolution is $1280 \times 960$ pixel$^2$ at a frame rate of 100 fps. Altogether, 8 different colors were available with 12 particles of each color. The remaining 144 non-tracked particles of each of the two types are black and serve as a `thermal' background. The choice of 96 colored particles in each experiment is approximately at the limits of the detection and tracking feasibility with the current setup and packing fraction. 

For particle detection and tracking, we followed the workflow which was described in \cite{Puzyrev2021}. The program was significantly improved, transferred to Detectron 2 framework \cite{Detectron2018}, and a custom GUI interface was added for preview and correction of 2D and 3D data.

A dataset which contains camera image frames together with the corresponding rod endpoints was assembled, first manually, later with an iterative procedure for corrected rods. The current dataset includes around 2500 images (around 300 000 object instances of colored rods) from both camera views from several independent runs of the experiments. Data processing scripts and pre-trained Detectron 2 network model files, as well as the GUI program for correction and annotation of data are being prepared for publication as an open-source software package.

In order to track the colored rods between the frames, rods endpoints are triangulated and matched by solving the 3D axial optimal assignment problem (also known as bipartite graph matching). We found that an optimization towards both reprojection error for rod endpoints and displacements of rod endpoints between frames is required for robust tracking of particles. 

The  trajectories of the rods were fitted taking into account constant translation velocities and rotation rates during the free flight phases of the particles between the collisions. For the rod centers of mass, the $l_1$ trend filtering optimization method \cite{Kim2009, Tibshirani2014} was used. It allows to fit the noisy particle center trajectories into a sequence of piecewise linear functions with kinks (bends of fitting function) in between. 

We extracted angular velocities by differentiating the rod orientation quaternions \cite{Shoemake1985,Huynh2009,RubioLargo2016}. To smoothen the high-frequency noise due to the orientation measurement error, a moving average filter was applied. Then, the resulting angular velocity data were fitted with the similar trend filtering approach, only instead of the piecewise linear approximation we fit the angular velocities with piecewise constant (step) functions.

For translational as well as angular velocities, we employ the special case the $l_1$ trend filtering for vector time series as described in \cite{Kim2009}. The advantage to fitting the $x,y,z$ center coordinates together is that the fitted coordinate components tends to show simultaneous trend changes at common kink points, which correspond to change of velocities due to collisions of the particle. One problem that arises when applying this procedure is that as particles slow down, the relative weight of the kinks in the optimization procedure decreases and the standard $l_1$ trend filtering procedure tends to overestimate the number of collisions. To improve the fitting, we have employed an iterative weighted heuristics from \cite{Kim2009} which allows to optimize the fit towards the number of kinks instead of sum of their residual norm. 

We assume that the optimization error which arises at the bending points of piecewise linear approximations of particle endpoint velocities, as well as jumps in angular velocities, signalize that the particle collided with another particle or the wall. Then, knowing the positions of the wall, collisions can be attributed to either particle-particle or particle-wall collision.  
This way, the collisions undergone by rods can be determined automatically, even though the absolute precision may not always be satisfactory.

\section*{Author contributions}
DP performed simulations, experimental data analysis and drafted the manuscript. All authors edited the manuscript. TT and DP performed drop tower experiments. TT, KH and RS prototyped, built and maintained experimental setup. All authors read and approved the final manuscript.

\section*{Acknowledgments}
The authors acknowledge funding by the German Aerospace Center (DLR) within grants 50W1842 and 50WM2048. We
cordially thank the ZARM Bremen staff for their support in the drop tower experiments, R.C. Hidalgo for providing the prototype simulation code and T. Steklova for the help with trajectory fitting. 

\section*{Competing interests}
All authors declare no financial or non-financial competing interests.

\section*{Data availability}
The datasets used and analysed during the current study are available from the corresponding author on reasonable request.

\section*{Code availability}
Particle detection and data correction code is publicly available at \url{https://github.com/ANP-Granular/ParticleTracking}. 
Other data processing scripts used in the current study are available from the corresponding author on reasonable request.

\end{document}